\documentclass[a4paper,12pt]{article}
\usepackage[T2A]{fontenc}
\usepackage[utf8]{inputenc}
\usepackage[english]{babel}
\usepackage{graphicx}
\usepackage{float}
\usepackage{caption}
\usepackage{amsmath}
\usepackage{amssymb}

\usepackage[a4paper,top=3cm,bottom=2cm,left=1.5cm,right=1.5cm,marginparwidth=0.05cm]{geometry}

\DeclareMathOperator{\sech}{sech}

\usepackage{authblk}

\title{Chaos in Coupled Heteroclinic Cycles between Weak Chimeras}
\author[1]{\small Artyom E. Emelin}
\author[1]{\small Evgeny A. Grines}
\author[1]{\small Tatiana A. Levanova}
\affil[1]{\footnotesize Lobachevsky University, Gagarin ave, 23, 603022 Nizhny Novgorod, Russia}
\affil[ ]{E-mail: \textbf{tatiana.levanova@itmm.unn.ru}}
\date{}

\begin{document}
\maketitle

\begin{abstract}
Heteroclinic cycles are widely used in neuroscience in order to mathematically describe different mechanisms of functioning of the brain and nervous system. Heteroclinic cycles and interactions between them can be a source of different types of nontrivial dynamics. For instance, as it was shown earlier, chaotic dynamics can appear as a result of interaction via diffusive couplings between two stable heteroclinic cycles between saddle equilibria. We go beyond these findings by considering two rotating in opposite directions coupled stable heteroclinic cycles between weak chimeras. Such an ensemble can be mathematically described by a system of six phase equations. Using two parameter bifurcation analysis we investigate the scenarios of appearance and destruction of chaotic dynamics in the system under study. 
\end{abstract}



\section*{INTRODUCTION}

One of the main goals of contemporary neurodynamics is to study the fundamental principles of the brain and nervous system functioning \cite{bick2020understanding}. Many mechanisms in neurodynamics and biology are based on various interacting periodic processes \cite{pikovsky2002synchronization}. In order to describe such processes one can use networks of different topologies with oscillators as network nodes \cite{strogatz2001exploring}. In this case, from the point of view of neurodynamics, a single element of such a network can describe a certain group of neurons in the nervous system \cite{breakspear2017dynamic}. 

The functioning of described networks highly depends on the properties and dynamics of its individual elements, as well as on the chosen topology and the strength of couplings \cite{winfree1980geometry}. For the interacting populations of identical oscillators a wide range of collective behaviour is possible, starting from global synchronization \cite{belykh2015dynamics,barabash2021partial} to the patterns of localized synchrony \cite{ashwin2015weak,bick2016chaotic,omel2018mathematics}. In the ensembles of coupled oscillators with higher-order interactions more sophisticated types of  dynamics are possible, e.g., heteroclinic switchings between patterns of localized frequency synchrony.

Studies of heteroclinic switchings between different metastable states in the context of computational neuroscience have a long history \cite{afraimovich2010robust,ashwin2011criteria,komarov2009sequentially,nekorkin2011relating,nekorkin2012reducing}. Starting from pioneering works by V. Afraimovich \cite{afraimovich2001chaotic,afraimovich2004origin,afraimovich2004heteroclinic}, where heteroclinic sequences consisting of saddles and heteroclinic orbits connecting them were shown to be a mathematical images of sequential activity in neural networks, to later research devoted to heteroclinic cycles based on various saddle limit sets, such as saddle cycles \cite{komarov2009sequentially,komarov2013heteroclinic,levanova2013sequential,mikhaylov2013sequential} and saddle chaotic sets \cite{dellnitz1995cycling,levanova2014coherence}, as well as to hierarchic heteroclinic structures of different topology \cite{afraimovich2014hierarchical,afraimovich2018mind}.

However, the description of metastable states is not limited to the saddle limit sets mentioned above. Chimera states, which consists of localized patterns of frequency synchronization, are also one of the possible candidates for describing such states. Dynamical transitions of the location of frequency synchrony were observed in \cite{bick2018heteroclinic}, where the system under study possessed metastable chimera states coupled by heteroclinic transitions. In subsequent papers \cite{bick2019heteroclinic} and \cite{bick2019heteroclinic_2}, the stability of heteroclinic cycles between chimera states was studied in detail. Also, previously obtained results were generalized to cases of a larger number of elements and different topologies of couplings between elements.

More complex structures  based on coupled subsystems, each of which possesses a stable heteroclinic cycle (here and further we will use the term "coupled heteroclinic cycles"), have attracted particular interest recently.  In \cite{li2012quasiperiodic} an example of a such system with nearly symmetric coupling between its subsystems was studied. Also, in the paper \cite{voit2020coupled} several identical heteroclinic cycles coupled by a diffusive term were studied. It was shown that in this case, with different parameters of the coupling strength, the system is able to reproduce various patterns of synchronous, quasi-periodic and chaotic dynamics. Subsequently, in the work \cite{pikovsky2023chaos} a system was proposed, consisting of two diffusively coupled subsystems, each of which is based on generalized a Lotka-Volterra model. It was shown that the described model demonstrates the predominant chaotic dynamics even for weak couplings between heteroclinic cycles.

The goal of this paper is to generalize this result to systems consisting of more complex heteroclinic cycles, in particular, to systems of coupled heteroclinic cycles between chimeric states. In order to do this we consider a system of six differential equations that describes the behavior of two interconnected clusters with oppositely directed heteroclinic cycles. Each cluster is described by a system of phase oscillators proposed in \cite{bick2019heteroclinic}. 
Based on the fundamental principles of constructing networks of phase oscillators, in our case, the most appropriate coupling function is selected, which is necessary to obtain various complex patterns of behavior of a dynamic system. Using various numerical methods, examples of the resulting collective dynamics will be demonstrated.

The paper is organized as follows. We introduce the model in Section \ref{sec:model} paying special attention to the following building blocks of the model. We discuss the basic heteroclinic cycle between chimera states in Subsection \ref{subsec:hc_btw_chimera}. We introduce couplings between two heteroclinic cycles and the resulting full model in Subsection \ref{subsec:hc_coupled}. The coupled system possesses several symmetries, which we discuss in Subsection \ref{subsec:symmetries}. 
Numerical exploration of described coupled heteroclinic cycles is performed in Section \ref{sec:results}. We discuss our findings and draw conclusion in  Section \ref{sec:conclusion}.

\section{THE MODEL}
\label{sec:model}

\subsection{Heteroclinic cycle between chimera states}
\label{subsec:hc_btw_chimera}

\begin{figure}[!tbh]
 \begin{center}
 \includegraphics[width=0.5\linewidth]{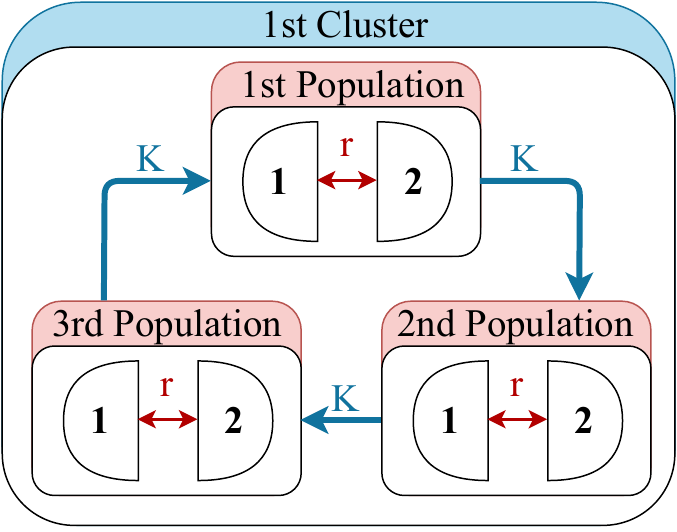}
 \caption{The topology of ensemble of phase oscillators that can reproduce heteroclinic cycle between chimera states.}
 \label{fig:single_ensemble}
 \end{center}
 \end{figure}

To describe the dynamics of one heteroclinic cycle, we will use the system of phase oscillators \cite{bick2019heteroclinic}. This model reproduces switching activity between chimera states in the sense of Ashwin-Burylko \cite{ashwin2015weak}. The model consists of three populations, each of which consist of two phase oscillators. The topology of described ensemble is presented in Fig. \ref{fig:single_ensemble}

Initially, in order to describe the dynamics of the heteroclinic cycle of this type one can use the following system of six ODEs:
\begin{equation}
\left\{\begin{array}{ll}
    \dot\theta_{\sigma,1} = \omega + g_2(\theta_{\sigma,2}-\theta_{\sigma,1})-K G_4(\theta_{\sigma-1};\theta_{\sigma,2} - \theta_{\sigma,1})+K G_4(\theta_{\sigma+1};\theta_{\sigma,2} - \theta_{\sigma,1})
\\
    \dot\theta_{\sigma,2} = \omega + g_2(\theta_{\sigma,1} - \theta_{\sigma,2})-K G_4(\theta_{\sigma-1};\theta_{\sigma,1} - \theta_{\sigma,2})+K G_4(\theta_{\sigma+1};\theta_{\sigma,1} - \theta_{\sigma,2})
\\
i \in \overline{1,6}
\end{array}\right. \label{eq:PhaseOsc}
\end{equation}
Here, interaction between oscillators within populations is pairwise and determined by the function
$$
g_2(\nu) = \sin(\nu + \alpha_2) - r \sin(2(\nu + \alpha_2)), 
$$ 
with parameters $\alpha_2$ and $r$. The interaction between populations is determined by the non-pairwise interaction function
$$
G_4(\theta_{\tau}; \phi) = \frac{1}{4} (g_4(\theta_{\tau, 1} - \theta_{\tau, 2} + \phi) + g_4(\theta_{\tau, 2} - \theta_{\tau, 1} + \phi)),
$$
where
$$
g_4(\nu) = \sin(\nu + \alpha_4).
$$
Here, the parameter $K$ is responsible for the strength of the connection between the populations. Parameter $\alpha_4$ is responsible for the interaction between populations.

Using reduction approach proposed in \cite{bick2019heteroclinic} we can reduce the dimension of the system to three equations. In order to do this we introduce a phase difference between phases of oscillators within each population, which allows us to rewrite the system \eqref{eq:PhaseOsc} in the following form:
\begin{equation}
\dot{\psi_{i}} = 
     \widehat{g_2}(\psi_{i})
    - \frac{K}{2} (\widehat{g_4}(\psi_{i-1} + \psi_{i}) + \widehat{g_4}(\psi_{i} - \psi_{i-1}))
    + \frac{K}{2} (\widehat{g_4}(\psi_{i+1} + \psi_{i}) + \widehat{g_4}(\psi_{i} - \psi_{i+1})),
    \label{Reduct_Sys}
\end{equation}
where
\begin{equation}
\widehat{g}_l(\nu) = \frac{1}{2}(g_l(-\nu) - g_l(\nu)), ~~ l \in \{2, 4\}
\end{equation}

\begin{figure}[!ht]
 \begin{center}
 (a)\includegraphics[width=0.45\linewidth]{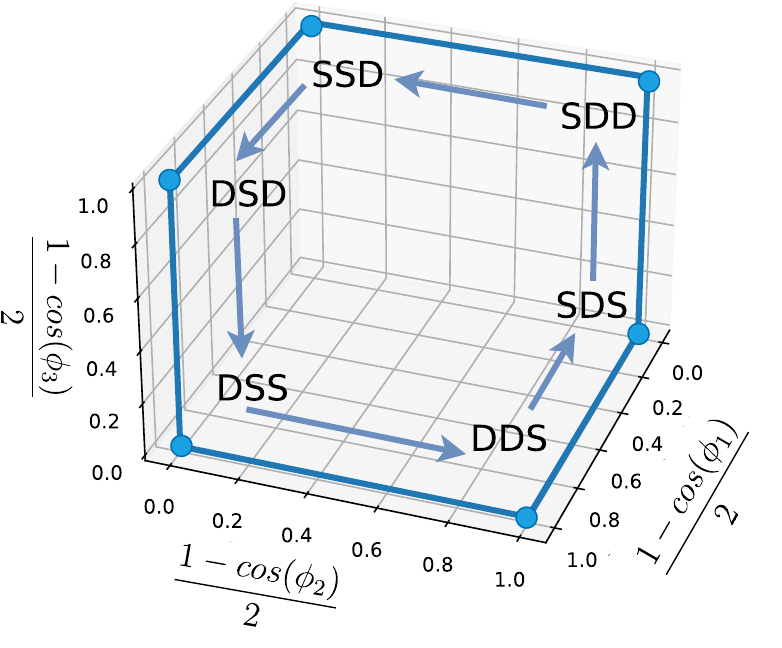}
 (b)\includegraphics[width=0.7\linewidth]{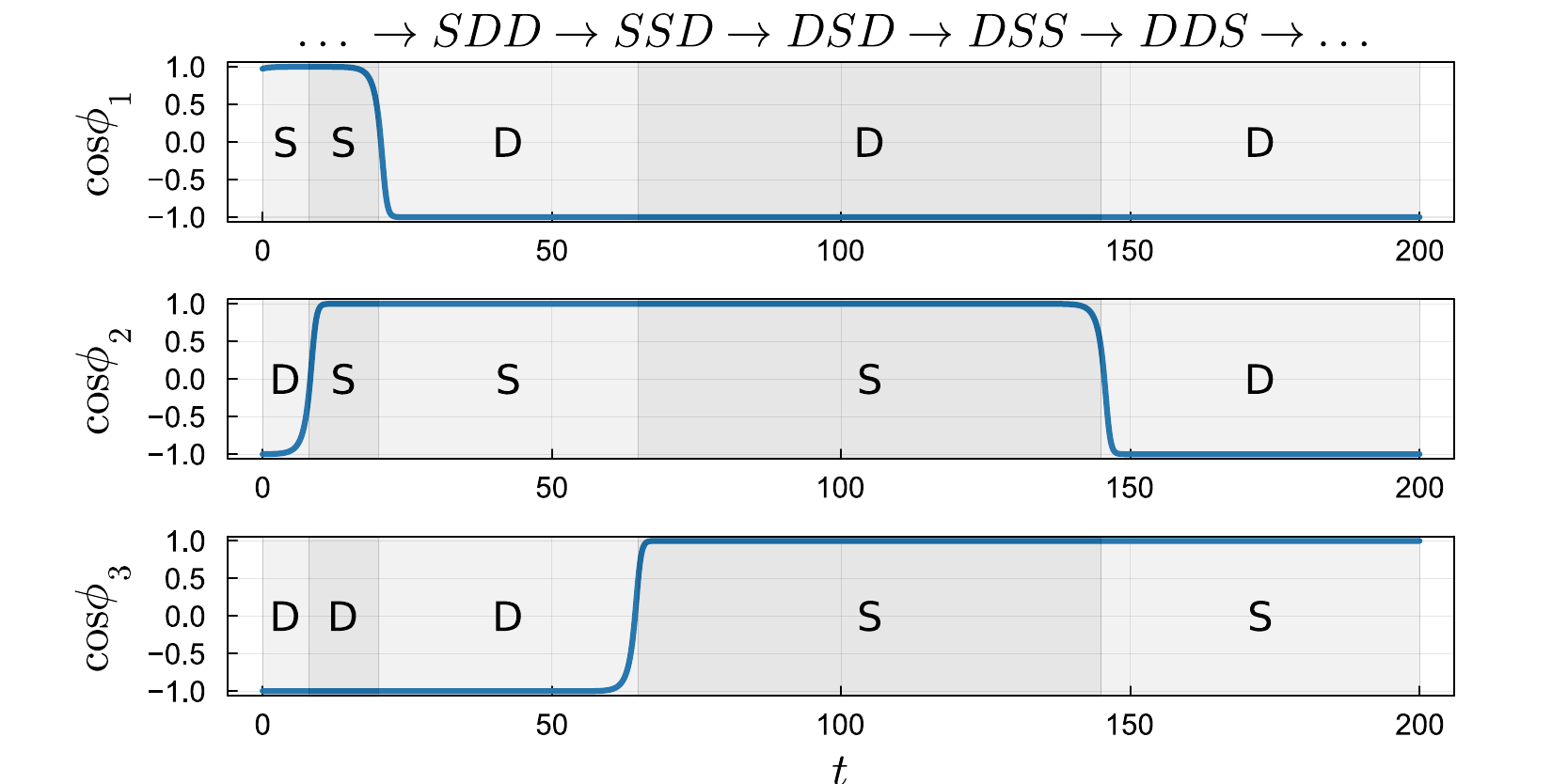}
 \caption{Sequence of metastable states in a system \eqref{Reduct_Sys}}
 \label{fig:DDD_SSS}
 \end{center}
 \end{figure}

It is easy to see that points of the phase space with any coordinate equal to 0 or $\pi$ are equilibrium states of the reduced system (\ref{Reduct_Sys}). Such equilibrium states have the following interpretation: the elements of the $i$-th population are in a synchronized state at the phase value $\phi_i = 0$ (we will denote it by the letter $S$), and a desynchronized state at the phase value $\phi_i = \pi$ (we will denote it by the letter $D$). All such combinations (except SSS and DDD) correspond to the saddle point equilibrium of the system (\ref{Reduct_Sys}). In \cite{bick2019heteroclinic} it was shown that in the system (\ref{Reduct_Sys}) there is a rough stable heteroclinic cycle that includes these metastable states (Fig. \ref{fig:DDD_SSS}). It is noteworthy that a change in the sign of the parameter $K$ leads to a change in the order of traversal of the heteroclinic cycle, but does not affect its stability. Clearly, if $(x(t), \, y(t), \, z(t))$ is a trajectory of system \eqref{Reduct_Sys} for a certain $K$, then it can be shown by direct calculations that $(x(t), \, z(t),\, y(t))$ is a solution of this system for $-K$ while the other parameters' values are fixed.

\subsection{Two coupled heteroclinic cycles}
\label{subsec:hc_coupled}

Let us start with two copies of system \eqref{eq:PhaseOsc} that differ from each other only in the sign of the parameter $K$. We will write it as follows:
\begin{equation}
\left\{\begin{array}{ll}
    \dot\theta_{\sigma,1} = \omega 
    + g_2(\theta_{\sigma,2}-\theta_{\sigma,1})
    -\,K G_4(\theta_{\sigma-1};\theta_{\sigma,2} - \theta_{\sigma,1})
    +\,K G_4(\theta_{\sigma+1};\theta_{\sigma,2} - \theta_{\sigma,1}) 
    +\,\varepsilon \mathcal{C}(\theta_{\sigma,2} - \theta_{\sigma,1}, \rho_{\sigma,2} - \rho_{\sigma,1}),
\\
    \dot\theta_{\sigma,2} = \omega 
    + g_2(\theta_{\sigma,1} - \theta_{\sigma,2})
    -\,K G_4(\theta_{\sigma-1};\theta_{\sigma,1} - \theta_{\sigma,2})
    +\,K G_4(\theta_{\sigma+1};\theta_{\sigma,1} - \theta_{\sigma,2}) 
    +\,\varepsilon \mathcal{C}(\theta_{\sigma,1} - \theta_{\sigma,2}, \rho_{\sigma,2} - \rho_{\sigma,1}),
\\
    \dot\rho_{\sigma,1} = \omega 
    + g_2(\rho_{\sigma,2}-\rho_{\sigma,1})
    +\,K G_4(\rho_{\sigma-1};\rho_{\sigma,2} - \rho_{\sigma,1})
    -\,K G_4(\rho_{\sigma+1};\rho_{\sigma,2} - \rho_{\sigma,1}) 
    +\,\varepsilon \mathcal{C}(\rho_{\sigma,2} - \rho_{\sigma,1}, \theta_{\sigma,2} - \theta_{\sigma,1}),
\\
    \dot\rho_{\sigma,2} = \omega 
    + g_2(\rho_{\sigma,1} - \rho_{\sigma,2})
    +\,K G_4(\rho_{\sigma-1};\rho_{\sigma,1} - \rho_{\sigma,2})
    -\,K G_4(\rho_{\sigma+1};\rho_{\sigma,1} - \rho_{\sigma,2}) 
    +\,\varepsilon \mathcal{C}(\rho_{\sigma,1} - \rho_{\sigma,2}, \theta_{\sigma,2} - \theta_{\sigma,1}),
\\
\sigma \in \overline{1,3}.
\end{array}\right. \label{eq:PhaseOscClusters}
\end{equation}

\begin{figure}[!tbh]
 \begin{center}
 \includegraphics[width=0.9\linewidth]{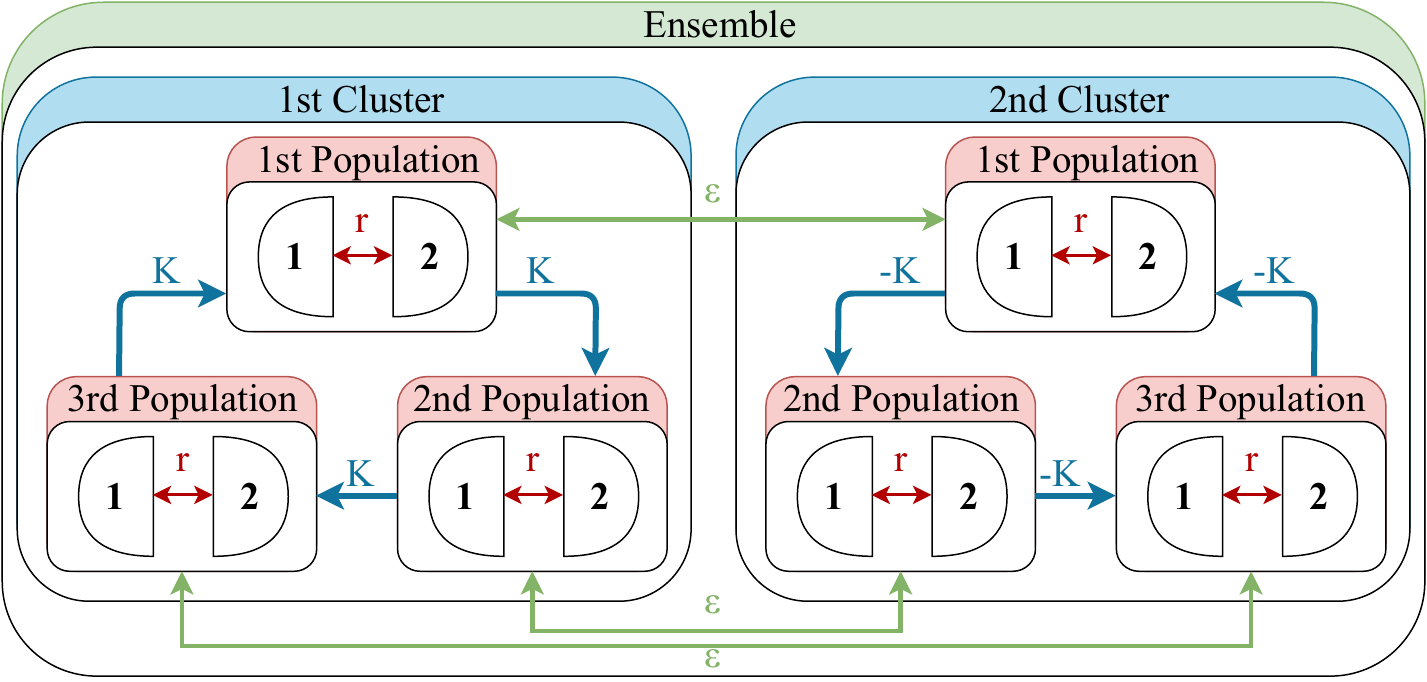}
 \caption{The topology of the system \eqref{eq:PhaseOscClusters}.}
 \label{fig:two_ensembles}
 \end{center}
 \end{figure}

We will call each of the groups of oscillators whose dynamics is described by the same letter a \textit{cluster}. The coupling function $\mathcal{C}(x, y)$ governs the interaction between populations of different clusters where $\varepsilon$ specifies the strength of this interaction. The topology of the resulting ensemble is presented in Fig. \ref{fig:two_ensembles}. 

Since equations \eqref{eq:PhaseOscClusters} still depend only on phase differences, a reduction similar to the one in the previous Subsection can be applied here. Namely, introducing variables 
$\varphi_\sigma = \theta_{\sigma, 2} - \theta_{\sigma, 1}$ and $\psi_\sigma = \rho_{\sigma, 2}  - \rho_{\sigma, 1}$ transforms equations \eqref{eq:PhaseOscClusters} into system 
\begin{equation}
    \left\{\begin{array}{l}
    \dot{\varphi}_{\sigma} = 
    2 \widehat{g_2}(\varphi_{\sigma}) 
    - \frac{K}{2} (\widehat{g_4}(\varphi_{\sigma-1} + \varphi_{\sigma}) + \widehat{g_4}(\varphi_{\sigma} - \varphi_{\sigma-1})) + ... \\
    ~~~~~~~~~~~~~~~~~~~~~~~~ ...+ \frac{K}{2} (\widehat{g_4}(\varphi_{\sigma+1} + \varphi_{\sigma}) + \widehat{g_4}(\varphi_{\sigma} - \varphi_{\sigma+1})) 
    + \varepsilon I(\varphi_{\sigma}, \psi_{\sigma}).

\\

    \dot{\psi}_{\sigma} = 
     2 \widehat{g_2}(\psi_{\sigma}) 
    + \frac{K}{2} (\widehat{g_4}(\psi_{\sigma-1} + \psi_{\sigma}) + \widehat{g_4}(\psi_{\sigma} - \psi_{\sigma-1})) - ... \\
    ~~~~~~~~~~~~~~~~~~~~~~~~ ...- \frac{K}{2} (\widehat{g_4}(\psi_{\sigma+1} + \psi_{\sigma}) + \widehat{g_4}(\psi_{\sigma} - \psi_{\sigma+1})) 
    + \varepsilon I(\psi_{\sigma}, \varphi_{\sigma}),  \\
\end{array}\right.
\label{eq:ensemble}
\end{equation}
where $I(x, y) = \mathcal{C}(-x, y) - \mathcal{C}(x, y)$ is a coupling function between variables describing the behaviour of populations in reduced system. 
While from the mathematical point of view the original inter-cluster coupling function $\mathcal{C}(x, y)$ can be any periodic function of its arguments, the inter-cluster coupling function between phase differences $I(x, y)$ clearly must be odd with respect to its first argument. 

Plugging the expressions for other coupling functions, we can rewrite Eqs. \eqref{eq:ensemble} as 
\begin{equation}
\label{eq:sysBasic}
\left\lbrace 
\begin{array}{ll}
\dot{\varphi}_1 = \left(-A \cos{\varphi_{2}} + A \cos{\varphi_{3}} + B \cos{\varphi_{1}} + C \right) \sin{\varphi_{1}} + \varepsilon I(\varphi_1, \psi_1),\\
\dot{\varphi}_2 = \left(-A \cos{\varphi_{3}} + A \cos{\varphi_{1}} + B \cos{\varphi_{2}} + C\right) \sin{\varphi_{2}} + \varepsilon I(\varphi_2, \psi_2),\\
\dot{\varphi}_3 = \left(-A \cos{\varphi_{1}} + A \cos{\varphi_{2}} + B \cos{\varphi_{3}} + C\right) \sin{\varphi_{3}} + \varepsilon I(\varphi_3, \psi_3),\\
\dot{\psi}_1 = \left(A \cos{\psi_{2}} - A \cos{\psi_{3}} + B \cos{\psi_{1}} + C \right) \sin{\psi_{1}} + \varepsilon I(\psi_1, \varphi_1),\\
\dot{\psi}_2 = \left(A \cos{\psi_{3}} - A \cos{\psi_{1}} + B \cos{\psi_{2}} + C\right) \sin{\psi_{2}} + \varepsilon I(\psi_2, \varphi_2),\\
\dot{\psi}_3 = \left(A \cos{\psi_{1}} - A \cos{\psi_{2}} + B \cos{\psi_{3}} + C\right) \sin{\psi_{3}} + \varepsilon I(\psi_3, \varphi_3),\\
\end{array}
\right . 
\end{equation}
where $A = K \cos{\alpha_4}$, $B = 4r \cos{2 \alpha_2}$ and $C = -2 \cos{\alpha_2}$. 

The system \eqref{eq:sysBasic} possesses an important property: $\varphi_i = \pi k$ and $\psi_i = \pi k$ are invariant hyperplanes for $k \in \mathbb{Z}$. In that case the phase space is divided into invariant cubes. 
For example, the cube $\lbrace \varphi_i \in \lbrack 0, \, \pi \rbrack, \; \psi_i \in \lbrack 0, \, \pi \rbrack \rbrace$ is invariant: its faces are invariant, and trajectories that start in the interior of this cube does not leave it due to the invariance of cubes' faces. 

\begin{figure}[!tbh]
(a)\includegraphics[width=0.45\linewidth]{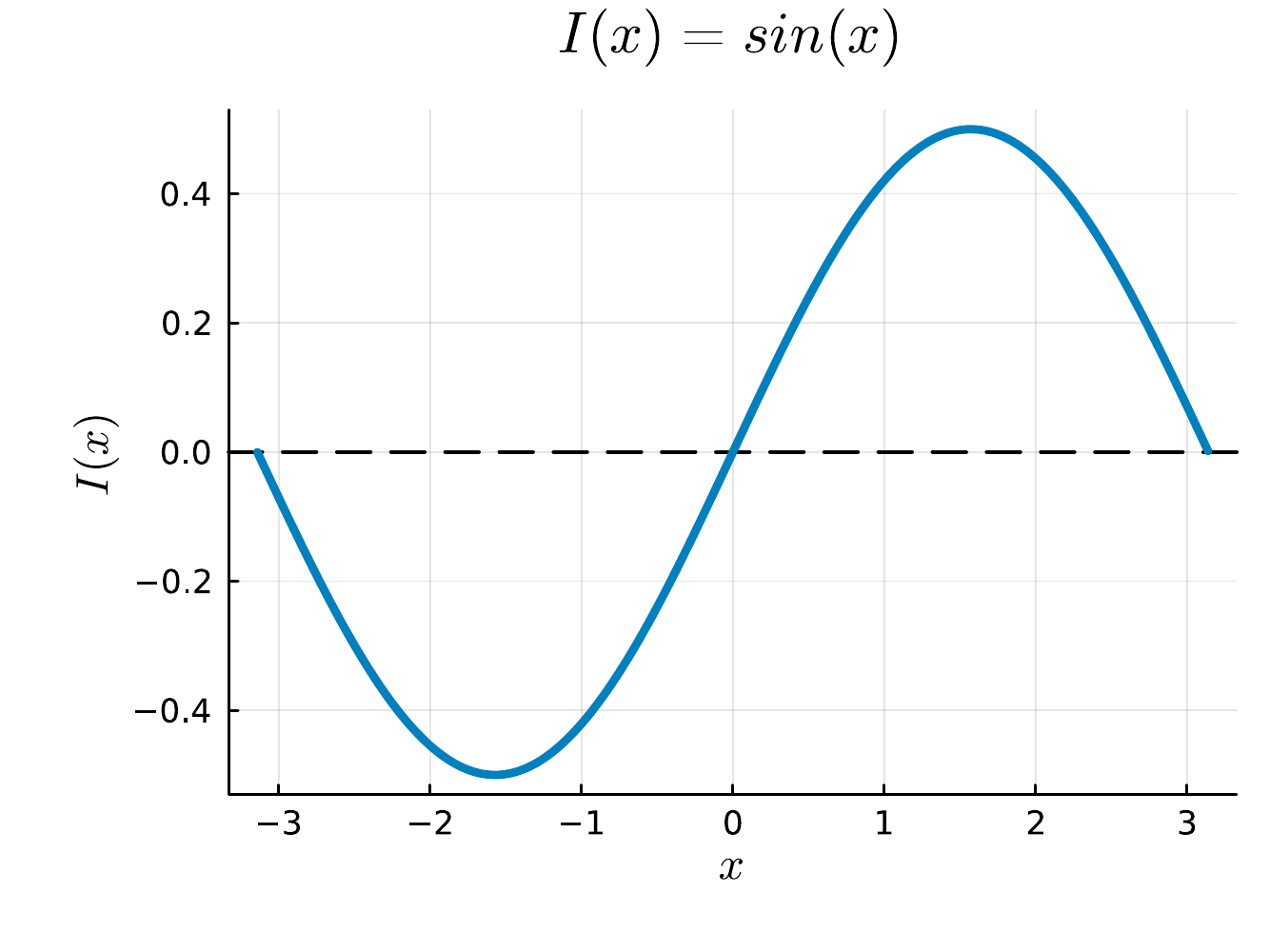}
(b)\includegraphics[width=0.45\linewidth]{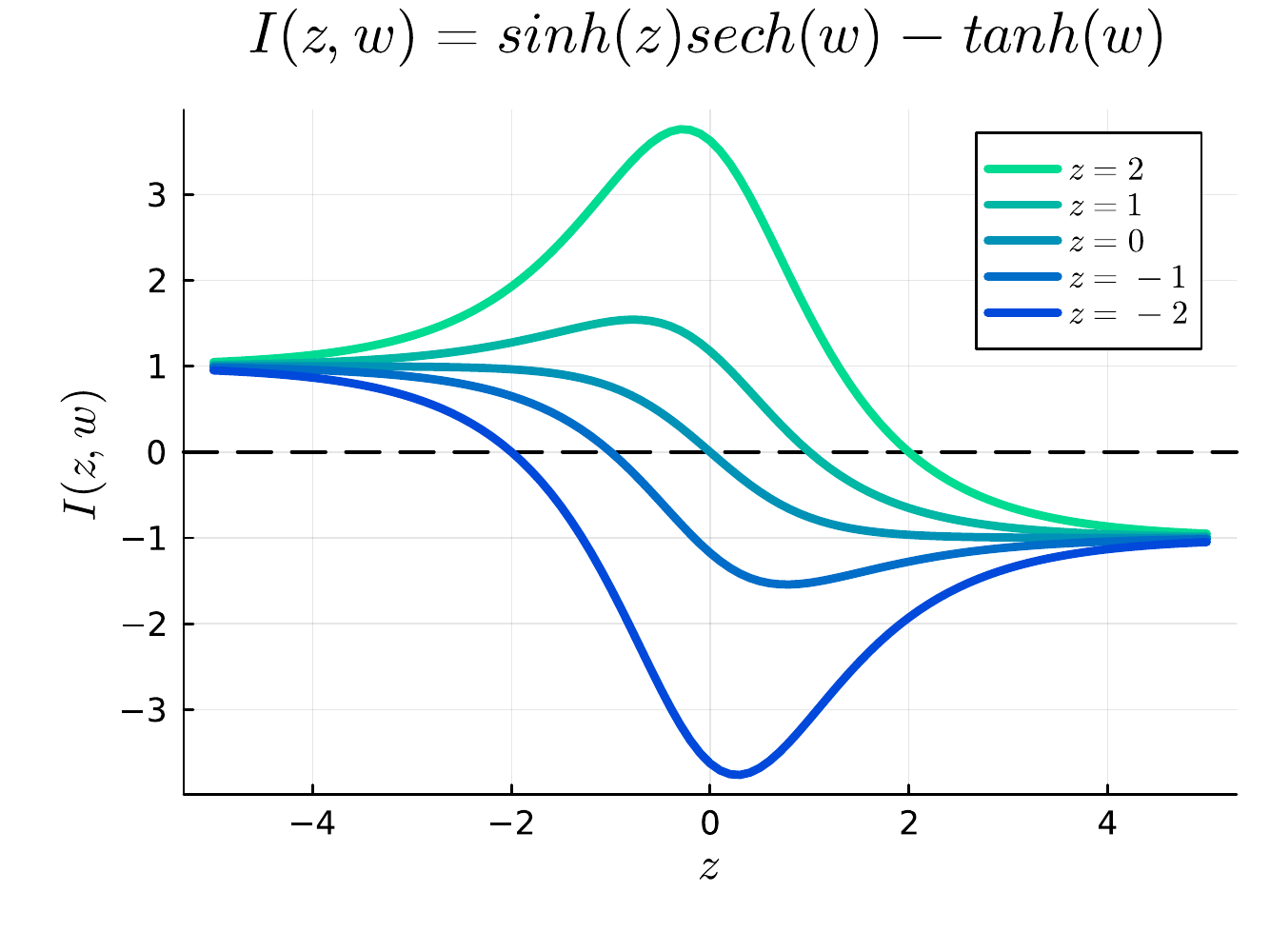}
\caption{Coupling functions. (a) $I(x) = \sin{x}$ and (b) $I(z, w) = \sinh{z} \sech{w} - \tanh{w}$}
\end{figure}

While this is true for ''natural'' coupling $I(x, y)$ that comes from full system \eqref{eq:PhaseOscClusters} with coupling $\mathcal{C}(x, y)$, some of these properties are preserved when we consider a "phase-like" system similar to Eqs.~\eqref{eq:sysBasic} where we do not require $I(x, y)$ to be odd in its first argument. 
Namely, from now on let us consider $$I(x, y) = \sin{(y-x)}.$$ 
Such choice breaks the invariance of cube's faces, but its interior is still invariant: the vector field at the boundary now points inside or it is tangent to it. Indeed, without loss of generality let us check $\dot{\varphi}_1$ when $\varphi_1 = 0$. 
Plugging that into Eqs.~\eqref{eq:sysBasic} gives $\dot{\varphi}_1 \vert_{\varphi_1 = 0} = \varepsilon \sin{\psi_1}$, which is non-negative when $\psi_1 \in \lbrack 0, \; \pi \rbrack$. 

When $\varphi_1 = \pi$, we obtain  $\dot{\varphi}_1 \vert_{\varphi_1 = \pi} = \varepsilon \sin{(\psi_1 - \pi)} = -\varepsilon \sin{\psi_1}$ which is non-positive when $\psi_1 \in \lbrack 0, \; \pi \rbrack$. Both these inequalities entail that vector field points inside of the cube at hyperplanes $\varphi_1 = 0$ and $\varphi_1 = \pi$. The same reasoning could be applied to all faces of the cube, hence we can conclude that trajectories starting in the cube's interior do not leave it. 

In order to study dynamics of system \eqref{eq:sysBasic} in this invariant cube we can utilize a coordinate transformation
\begin{equation}
\label{eq:tanSubst}
z_k = \mathcal{F}(\varphi_k), \; 
w_k = \mathcal{F}(\psi_k), 
\end{equation}
where 
\begin{equation}
\label{eq:tanFormula}
\mathcal{F}(x) =  \ln \sqrt{\frac{1 - \cos{x}}{1 + \cos{x}}} = \ln \tan{\frac{x}{2}}.
\end{equation}
Such transformation is a monotonous map from the interval $(0, \pi)$ to the interval $(-\infty, +\infty)$: the values close to $0$ are mapped close to $-\infty$, whereas values close to $\pi$ are mapped close to $+\infty$. 
This transformation is inspired by the work of Pikovsky and Nepomnyaschy \cite{pikovsky2023chaos} and serves the same purpose.  In the absence of coupling (i.e., when $\varepsilon = 0$) the attractors of the system are likely to be situated on the invariant faces of the cube. 
In that case a simulation of dynamics might become very sensitive to unavoidable errors of approximation: numerical trajectories might either stuck close to saddle equilibria or step outside of invariant cube due to imprecise calculations. 
This transformation plays a role of a "magnifying glass": when attractors are situated on the cube's faces, variables $z_k$ and $w_k$ tend to infinity, whereas they stay finite even when an attractor is close to the previously invariant faces. 

Let us write a transformed version of Eqs.~\eqref{eq:sysBasic} using the substitution \eqref{eq:tanSubst}. 
Using tangent half-angle substitution
$$
\sin{x} = \frac{2 u}{1 + u^2}, \; \cos{x} = \frac{1 - u^2}{1 + u^2}, \; u = \tan{\frac{x}{2}},
$$
and utilizing reordered version of relation between angular and logarithmic coordinates 
$$
e^{{z_k}} = \tan{\frac{\varphi_k}{2}}, \; 
e^{{w_k}} = \tan{\frac{\psi_k}{2}},
$$
we can express basic trigonometric functions of $\varphi_k$ and $\psi_k$ in terms of hyperbolic functions of $z_k$ and $w_k$:
\begin{equation}
\label{eq:trighyper}
\cos{\varphi_k} = \frac{1 - \tan^2{\frac{\varphi_k}{2}}}{1 + \tan^2{\frac{\varphi_k}{2}}} = \frac{1 - e^{2z_k}}{1 + e^{2z_k}} =  -\tanh{{z_k}}, \; 
\sin{\varphi_k} = \frac{2 \tan{\frac{\varphi_k}{2}}}{1 + \tan^2{\frac{\varphi_k}{2}}} =  \frac{2 e^{{z_k}}}{1 + e^{2z_k}} = \frac{1}{\cosh{{z_k}}};    
\end{equation}
similar formulas can be written for $\psi_k$ and $w_k$.

Finally, combining formulas \eqref{eq:trighyper} and the fact that $\mathcal{F}'(x) = \frac{1}{\sin{x}}$, we arrive at the transformed system of equations in logarithmic coordinates:
\begin{equation}
\label{eq:logSys}
\left \lbrace
\begin{array}{ll}
\dot{z}_1 = \left(A \tanh{{z_2}} - A \tanh{{z_3}} - B \tanh{{z_1}} + C \right)  + \varepsilon \bigl ( \tanh{{w_1}} - \sinh{{z_1}}\; \sech{{w_1}}  \bigr ),\\
\dot{z}_2 = \left(A \tanh{{z_3}} - A \tanh{{z_1}} - B \tanh{{z_2}} + C \right)  + \varepsilon \bigl ( \tanh{{w_2}} - \sinh{{z_2}}\; \sech{{w_2}}  \bigr ),\\
\dot{z}_3 = \left(A \tanh{{z_1}} - A \tanh{{z_2}} - B \tanh{{z_3}} + C \right)  + \varepsilon \bigl ( \tanh{{w_3}} - \sinh{{z_3}}\; \sech{{w_3}}  \bigr ),\\
\dot{w}_1 = \left(-A \tanh{{w_2}} + A \tanh{{w_3}} - B \tanh{{w_1}} + C \right)  + \varepsilon \bigl ( \tanh{{z_1}} - \sinh{{w_1}}\; \sech{{z_1}}  \bigr ),\\
\dot{w}_2 = \left(-A \tanh{{w_3}} + A \tanh{{w_1}} - B \tanh{{w_2}} + C \right)  + \varepsilon \bigl ( \tanh{{z_2}} - \sinh{{w_2}}\; \sech{{z_2}}  \bigr ),\\
\dot{w}_3 = \left(-A \tanh{{w_1}} + A \tanh{{w_2}} - B \tanh{{w_3}} + C \right)  + \varepsilon \bigl ( \tanh{{z_3}} - \sinh{{w_3}}\; \sech{{z_3}}  \bigr ).
\end{array}
\right .
\end{equation}
While all terms involving $\tanh$ and $\sech$ functions are bounded, the only $\sinh$ term in each equation grows unbounded and counteracts an unlimited growth of phase variable to either of infinities (at least, theoretically).

\subsection{Symmetries}
\label{subsec:symmetries}

Let us highlight the symmetry properties of the system \eqref{eq:logSys}. 
It is well known that a presence of symmetries influences both possible geometries of attractors (which can be asymmetric or posess some subgroup of self-symmetries) and their possible bifurcations. This short section forms the basis for an explanation in Section \ref{sec:results} of pitchfork bifurcation curves appearing in parameter space. 

Recall that a transformation of a phase space $\mathcal{T}: \mathbb{R}^n \mapsto \mathbb{R}^n$ is called a \textit{symmetry} of an ODE $\dot{x} = F(x)$ if for any solution $\gamma(t)$ of this equation follows that $\mathcal{T}(\gamma(t))$ is also a solution of this ODE. 
From this definition follows a computable criterion: a map $\mathcal{T}$ is a symmetry if $F(\mathcal{T}(x)) \equiv \mathcal{T}'(x) \cdot F(x)$ for any $x \in \mathbb{R}^n$, where $\mathcal{T}'(x)$ is a Jacobi matrix of the map $\mathcal{T}$ computed at the point $x$. 
The symmetries of \eqref{eq:logSys} are the same as in \cite{pikovsky2023chaos} due to the same coupling principle between clusters and the same symmetries of original equations \eqref{eq:PhaseOsc} for a single cluster. 

Using the notation from \cite{pikovsky2023chaos}, we can say that system \eqref{eq:logSys} is symmetric with respect to following transformations:
\begin{itemize}
\item {renamings of variables inside each subsystem:
$$
\begin{aligned}
& \quad R_1:\left(z_1, z_2, z_3, w_1, w_2, w_3\right) \rightarrow\left(z_2, z_3, z_1, w_2, w_3, w_1\right) \\
& \text { and } R_2:\left(z_1, z_2, z_3, w_1, w_2, w_3\right) \rightarrow\left(z_3, z_1, z_2, w_3, w_1, w_2\right) \text {. }
\end{aligned}
$$}
\item{exchange transformations between subsystems:
$$
\begin{gathered}
T_1:\left(z_1, z_2, z_3, w_1, w_2, w_3\right) \rightarrow\left(w_1, w_3, w_2, z_1, z_3, z_2\right), \\
T_2:\left(z_1, z_2, z_3, w_1, w_2, w_3\right) \rightarrow\left(w_3, w_2, w_1, z_3, z_2, z_1\right), \\
\text { and } T_3:\left(z_1, z_2, z_3, w_1, w_2, w_3\right) \rightarrow\left(w_2, w_1, w_3, z_2, z_1, z_3\right) .
\end{gathered}
$$} 
\end{itemize}
There are following relations between these symmetries: $R_2=R_1^{-1}=R_1^2$, $R_1=R_2^{-1}=R_2^2$, $T_1^2=T_2^2=T_3^2=I$,$ T_2=R_2 T_1$, $T_3=R_1 T_1$, where $I$ is an identity map.

\section{THE RESULTS}
\label{sec:results}

In order to investigate the phenomenon of appearance of chaotic dynamics in the system \eqref{eq:logSys} we conducted a numerical study. The results discussed below were obtained by us using Julia software \cite{datserisdynamicalsystems} and a package for numerical bifurcation analysis MATCONT \cite{MATCONT}.

\begin{figure}[!tbh]
\includegraphics[width=0.9\linewidth]{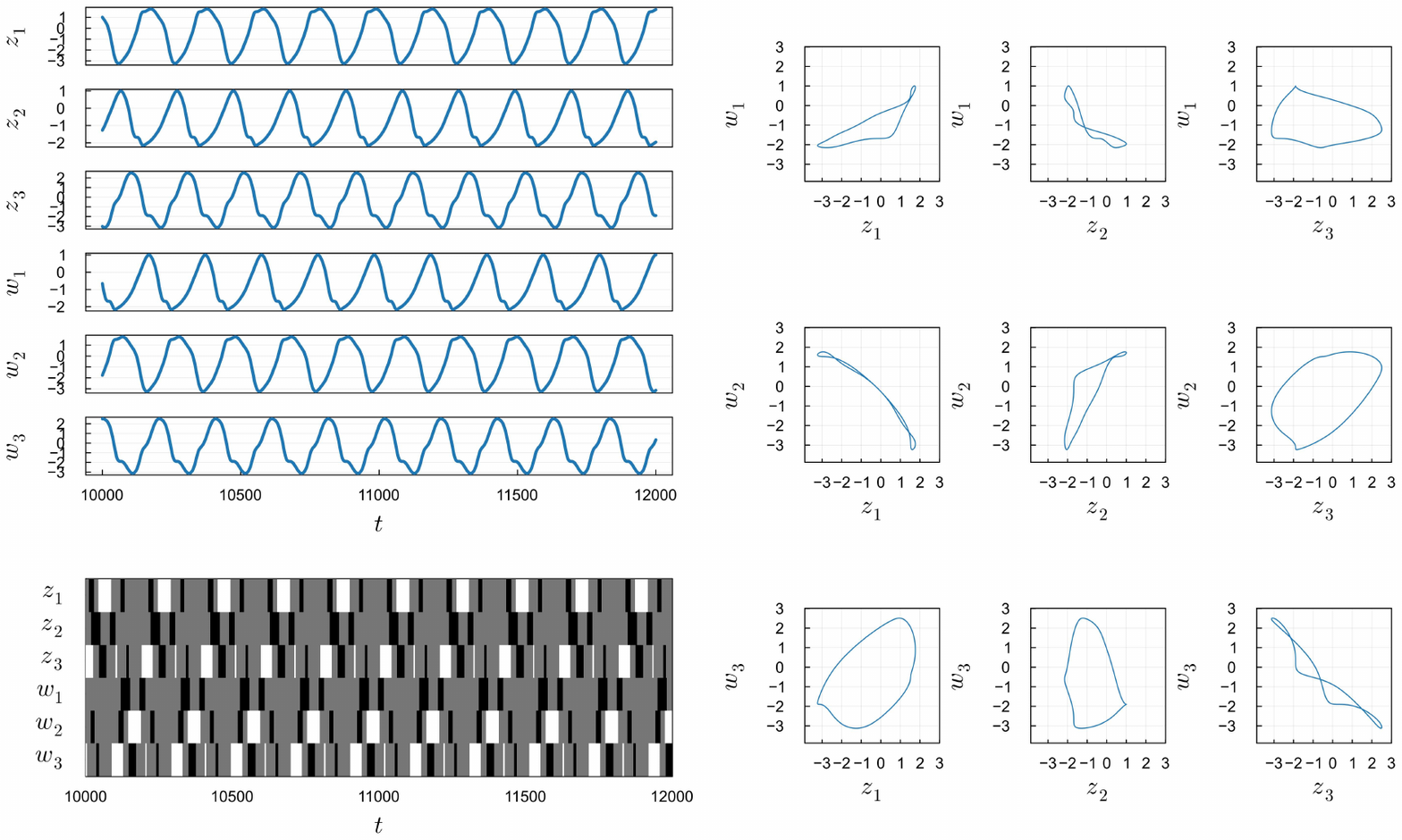}
\caption{Regular dynamics in a system \eqref{eq:logSys}. Left panel: time series. Right panel: projections of 6-dimensional phase space of the system \eqref{eq:logSys} on the different phase planes. Parameter values: $\varepsilon=0.05206157664979711$, $K=0.05977203237457928$.}
\label{fig:regular_dynamics}
\end{figure}

Our starting point for studying and describing dynamics of the system \eqref{eq:logSys} is a stable limit cycle found at parameter values
$$\begin{array}{c} 
\alpha_2 = \frac{\pi}{2}, \; r = 0.016850701746869687, \; \alpha_4 = \pi, \\K = 0.05977203237457928, \varepsilon = 0.05206157664979711.
\end{array}
$$
This is a comparatively large-period limit cycle (the period is roughly $200$ time units according to MATCONT), see Fig.~\ref{fig:regular_dynamics}. 
This limit cycle $\mathcal{L}$ was used as a starting point for an attractor continuation in parameters $\varepsilon$ and $K$, leaving all other parameter values fixed. 


\begin{figure}[!tbh]
\includegraphics[width=1\linewidth]{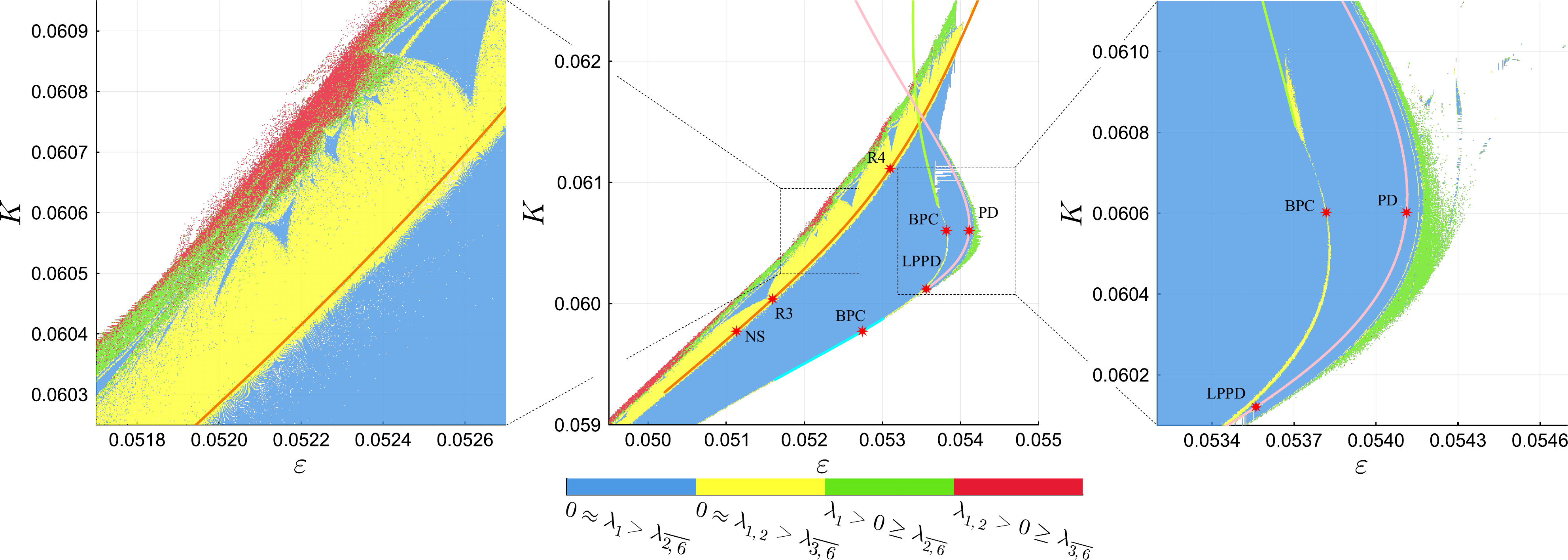}
\caption{The map of the Lyapunov spectrum of the system \eqref{eq:logSys} and its enlarged fragments. See detailed legend in subfigures}
\label{fig:spectrum_map}
\end{figure}

A map of the Lyapunov spectrum based on this attractor continuation is presented on Fig.~\ref{fig:spectrum_map}. 
Here the red marker corresponds to the regions with hyperchaotic dynamics (i.e. regions with two positive Lyapunov exponents, $\lambda_1>\lambda_2>0$), the green marker -- to the regions with chaotic dynamics (one positive Lyapunov exponents, $\lambda_1>0$), yellow marker -- to the regions with torus attractors ($\lambda_{1, 2} \approx 0 > \lambda_{3, 4, 5, 6}$), blue marker -- to the regions with limit cycle attractors ($\lambda_1 \approx 0 > \lambda_{2, 3, 4, 5, 6}$), and white marker -- to the regions where trajectories of the system \eqref{eq:logSys} diverge to infinity. A detailed description of the calculation of maps of the largest Lyapunov exponent can be found, e.g., in the following studies \cite{borisov2012dynamical, borisov2014reversal}.


While theoretically equations \eqref{eq:logSys} preclude unbounded growth of solutions, in certain cases such effect occurs even during attractor continuation. Here we consider a trajectory going to infinity if any of the phase variable exceeds 20 in absolute value. It corresponds to a trajectory of the system \eqref{eq:sysBasic} being at distance of $4.12 \times 10^{-9}$ (an approximate value of $2 \cdot \arctan{e^{-20}}$, see formula \eqref{eq:tanFormula}) to $0$ or ${\pi}$ in some of its phase variable. 
Let us describe some features of dynamics in the region of bounded dynamics.

Continuing the limit cycle $\mathcal{L}$ with fixed $K$ and decreasing $\varepsilon$ leads to its loss of stability in a Neimark-Sacker bifurcation.
This observation agrees with MATCONT computations, which locate the position of $R_3$ and $R_4$ bifurcation points at the tips of periodicity windows found at the map of Lyapunov exponents.
Further change of parameters to the left of Neimark-Sacker curve leads to a chaotization of attracting invariant tori and formation of chaotic and hyperchaotic regimes. 

\begin{figure}[!tbh]
\includegraphics[width=0.9\linewidth]{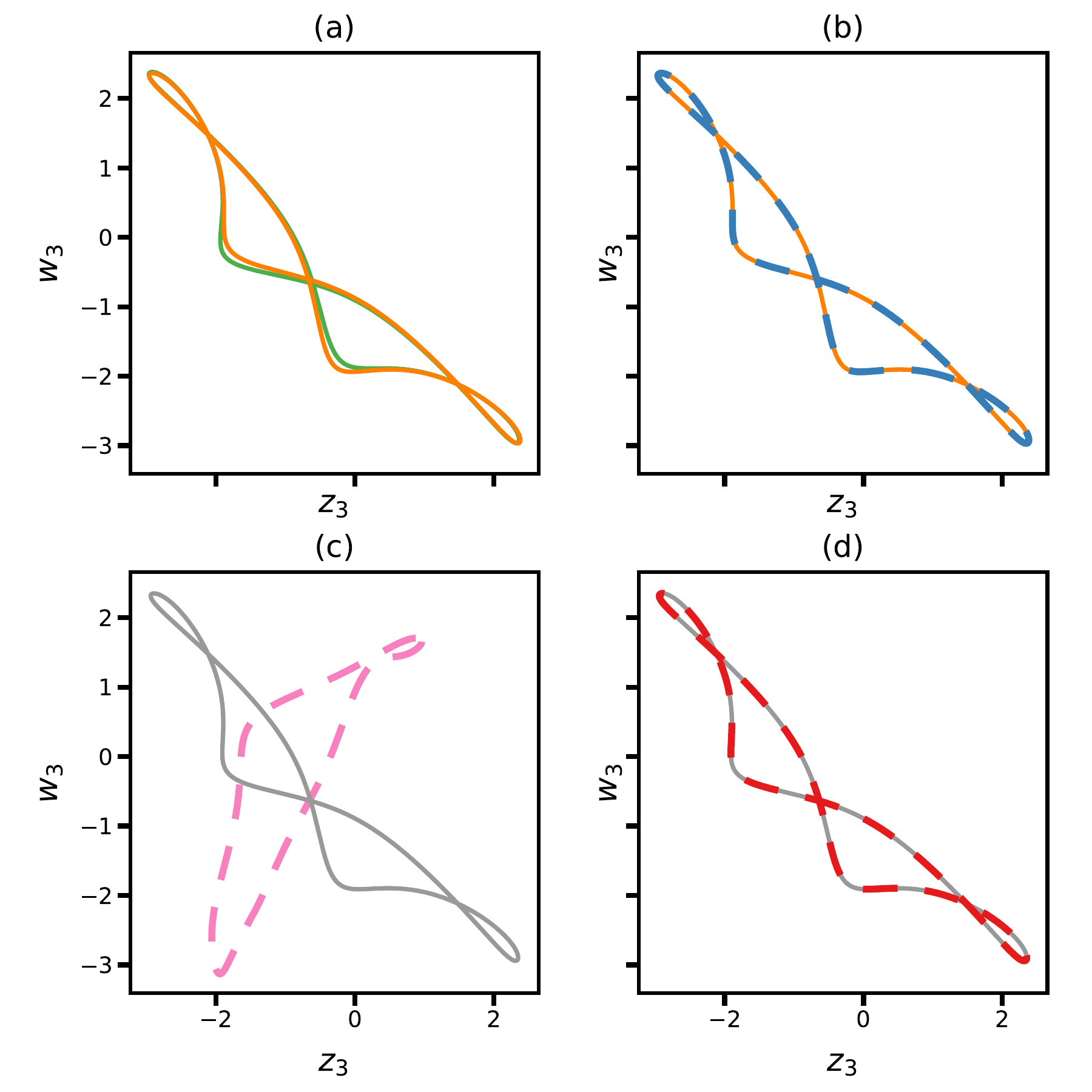}
\caption{A setup for a pitchfork bifurcation, $K=0.05977203237457928$. Two unstable limit cycles $\mathcal{C}_1$ and $\mathcal{C}_2$ collide with a stable limit cycle at $\varepsilon = 0.05274407266311118 $, which leads to its loss of stability. (a) Projections of cycles $\mathcal{C}_1$ (green) and $\mathcal{C}_2$ (orange) show that they are visibly different at $\varepsilon \approx 0.0527226 $ before the bifurcation; (b) however, they are symmetry related: $\mathcal{C}_2$ (orange) and image of $\mathcal{C}_1$ (blue) under symmetry $T_3$ are the same set-wise; (c) limit cycle $\mathcal{C}$ (gray) and its $T_2$ image (pink) are visibly different; (d) meanwhile, $\mathcal{C}$ (gray) and its $T_3$ image (red) are the same set-wise.}
\label{fig:beforeBPC}
\end{figure}

Continuing the limit cycle $\mathcal{L}$ to the right by increasing $\varepsilon$ leads to the loss of its stability in a subcritical pitchfork bifurcation. 
While in generic systems pitchfork bifurcation is usually a codimension-2 phenomenon, here it is a codimension-1 bifurcation due to the presence of symmetries. 
Fig.~\ref{fig:beforeBPC} illustrates the details of this bifurcation. 
Here the initial stable limit cycle $\mathcal{L}$ already has a non-trivial \textit{isotropy subgroup}, i.e. a subgroup of symmetry group that leaves limit cycle $\mathcal{L}$ invariant as a set. 
Namely, this limit cycle is mapped to itself by a symmetry $T_3$ as illustrated on Fig.~\ref{fig:beforeBPC}d, while other symmetries map it to a completely different set (as an example, action of $T_2$ on it is illustrated on Fig.~\ref{fig:beforeBPC}c). 

However, the unstable limit cycles $\mathcal{C}_1$ and $\mathcal{C}_2$ that exist before bifurcation happens do not posess this self-symmetry, but are mapped to each other by this $T_3$ symmetry.
As parameter changes, they both get close to a limit cycle $\mathcal{L}$ and collide with it, leading to a stability loss; thus, a subcritical pitchfork bifurcation due to the system being equivariant.
After the loss of stability the trajectories go to a new attractor that is not located near the old one. 
This repeats in other parts of parameter space, which might explain a certain artifacts of the map of Lyapunov exponents: in that case attractor and its characteristics might jump abruptly. 

A presence of multiple coexisting attractors can be further illustrated by the right insert on Fig.~\ref{fig:spectrum_map}.
Here an attracting limit cycle found in that region was used for numerical continuation in MATCONT. 
Fixing $K$ as in previous case and varying $\varepsilon$ we find two bifurcation curves that bound stability region of this regime: a curve of pitchfork bifurcation and a curve of period-doubling bifurcation.
The right insert also shows other bifurcations that happen during this continuation, revealing a complex organization of parameter and phase space.
While we were unsuccessful in further numerical continuation of large-period limit cycles (period is $\approx 400$) and discovering other period doubling curves, we expect that in this region chaotic dynamics occurs through a cascade of period doubling bifurcations.

\begin{figure}[!tbh]
\includegraphics[width=0.9\linewidth]{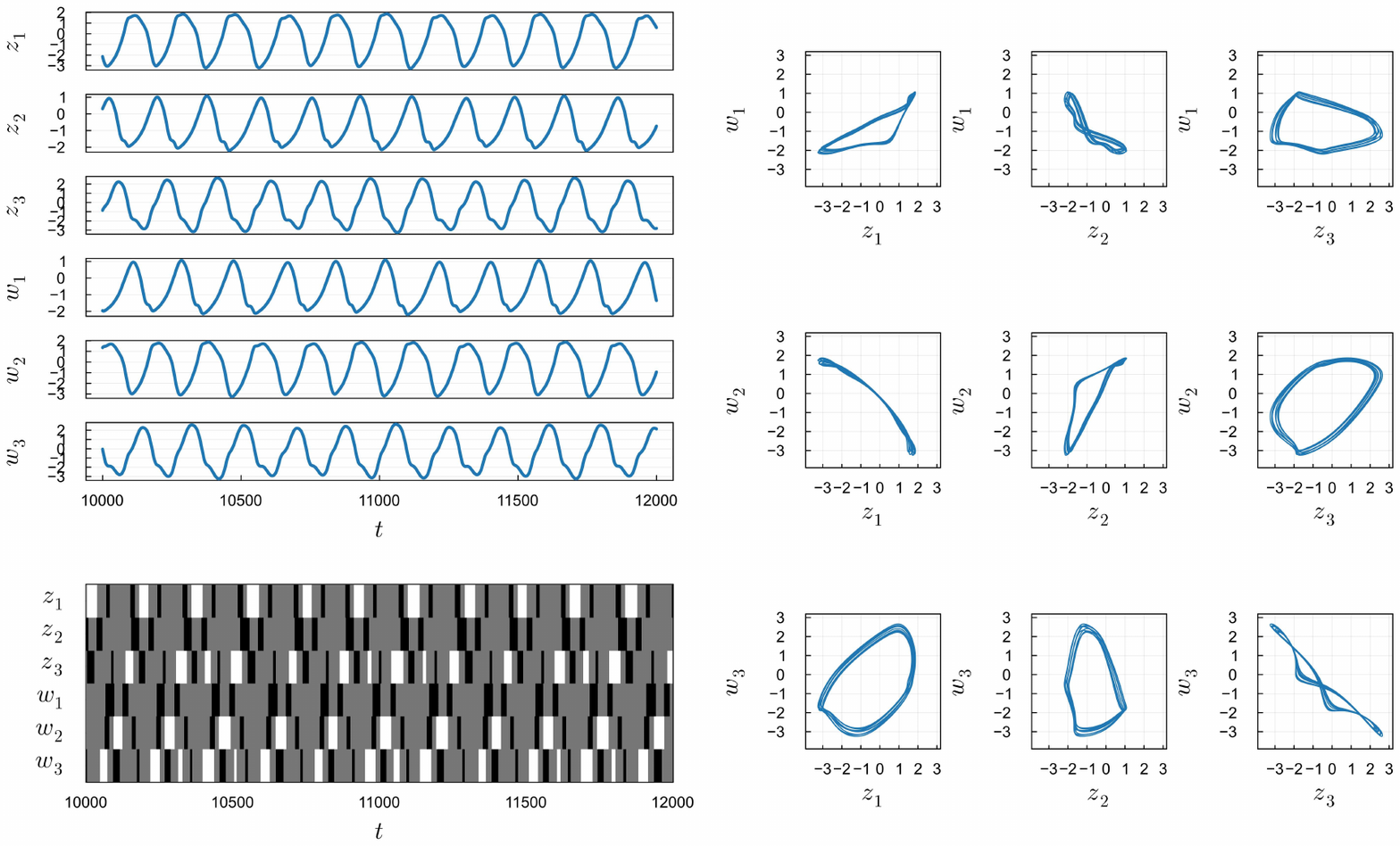}
\caption{Chaotic dynamics in a system \eqref{eq:logSys}. Left panel: time series. Right panel: projections of 6-dimensional phase space of the system \eqref{eq:logSys} on the different phase planes. Parameter values: $\varepsilon=0.0527$, $K=0.06096$.}
\label{fig:chaotic_dynamics}
\end{figure}

Let us describe numerically typical dynamical regimes that can be observed in the system \eqref{eq:logSys} for different combinations of governing parameters $\varepsilon$ and $K$ in the Fig. \ref{fig:spectrum_map}. Examples of regular and chaotic dynamics are presented in Fig. \ref{fig:regular_dynamics} and Fig. \ref{fig:chaotic_dynamics}, correspondingly. 

\section{CONCLUSION}
\label{sec:conclusion}

In this paper, we have studied coupled heteroclinic cycles between chimera states rotating in opposite directions on the basis of the system of phase equations. We generalize earlier result by Pikovsky and Nepomnyashchy \cite{pikovsky2023chaos}, where the appearance of chaotic dynamics was shown as a result of interaction between oppositely rotating heteroclinic cycles between saddle equilibria for the case of weak diffusive coupling between heteroclinics. Using two-parameter bifurcation analysis we have identified regions on the plane of governing parameters where different types of nontrivial dynamics exist, including hyperchaotic, chaotic and regular spatio-temporal patterns. Note that in our numerical simulations of the system under study, we have observed that chaotic and hyperchaotic dynamics arise for small values of inter-cluster and intra-cluster coupling strengths. The increase in these coupling strength leads to a series of bifurcation transitions related to the change of symmetry of the chaotic attractors. 

We assume that the origin of chaotic and hyperchaotic dynamics in similar systems is a general effect that can be observed in a range of governing parameters for heteroclinic cycles between different types of attracting sets (e.g. saddle cycles, saddle chaotic sets etc.) and various types of couplings between them (e.g. mutual synaptic couplings). Nevertheless, the verification of this hypothesis is a subject of future studies. 


This paper is devoted to Prof. V. N. Belykh, who contributed greatly to the various field of nonlinear dynamics, such as studies of synchronization and clustering in ensembles of coupled dynamic systems \cite{belykh2004connection,belykh2004blinking,barabash2018synchronization,belykh2016bistability,brister2020three,barabash2021partial}, existence of homoclinic \cite{belykh1997homoclinic,belykh2010chaotic,belykh1984bifurcation} and heteroclinic orbits and neurodynamics \cite{belykh2000homoclinic,belykh2005hyperbolic,belykh2008dynamics,barabash2020ghost}. 




 \section*{ACKNOWLEDGMENTS}

 The authors thank Prof. A. S. Pikovsky and Prof. G.V. Osipov for useful discussions.

 \section*{FUNDING}

 This work was supported by the Ministry of Science and Education of Russian Federation Contract no. FSRW-2020-0036 (A.E.E. and E.A.G.) and RSF grant 22-12-00348 (T.A.L.).

 \section*{CONFLICT OF INTEREST}

 The authors declare that they have no conflicts of interest.




\bibliographystyle{unsrt}
\bibliography{RCD_arxiv}


\end{document}